\documentclass[epj]{svjourmod}

\usepackage{amsmath,amssymb,amsfonts}
\usepackage{graphicx}
\usepackage{citesort}

\newcommand{\beq}{\begin{equation}}
\newcommand{\eeq}{\end{equation}}
\newcommand{\Order}{\mathcal{O}}
\newcommand{\Lagr}{\mathcal{L}}
\newcommand{\M}{\mathcal{M}}
\newcommand{\p}{\mathcal{P}}
\newcommand{\N}{\mathcal{N}}

\newcommand{\mep}{M_{\eta'}^2}
\newcommand{\me}{M_\eta^2}
\newcommand{\mpi}{M_\pi^2}
\newcommand{\mpn}{M_{\pi^0}^2}
\newcommand{\Fpi}{F_\pi}
\newcommand{\mk}{M_K^2}

\newcommand{\Amp}{\mathcal{M}}
\newcommand{\nn}{\nonumber\\}
\newcommand{\logpi}{\log\frac{M_\pi}{\mu}}
\newcommand{\logk}{\log\frac{M_K}{\mu}}
\newcommand{\loge}{\log\frac{M_\eta}{\mu}}
\newcommand{\mc}{\multicolumn{3}{c}}
\newcommand{\tbk}{\hspace{-3.5mm}}
\newcommand{\tpm}{\tbk$\pm$}

\begin{document}

\title{\\[-0.95cm]\phantom{ }\hfill{\scriptsize \textnormal{HISKP--TH--09/13}}\\[1mm]
The cusp effect in \boldmath{$\eta'\to\eta\pi\pi$} decays}
\titlerunning{The cusp effect in $\eta'\to\eta\pi\pi$ decays}

\author{Bastian Kubis, Sebastian P. Schneider}

\institute{
   Helmholtz-Institut f\"ur Strahlen- und Kernphysik (Theorie)
   and 
   Bethe Center for Theoretical Physics,
   Universit\"at Bonn, \linebreak D--53115~Bonn, Germany
}

\authorrunning{B. Kubis and S. P. Schneider}

\date{
}

\abstract{
Strong final-state interactions create a pronounced cusp in $\eta'\to\eta\pi^0\pi^0$ decays.
We adapt and generalize the non-relativistic effective field theory framework developed for the extraction of 
$\pi\pi$ scattering lengths from $K\to3\pi$ decays to this case.
The cusp effect is predicted to have an effect of more than 8\% on the decay spectrum
below the $\pi^+\pi^-$ threshold.
}

\maketitle

\section{Introduction}

\begin{sloppypar}
In the last few years, the investigation of the cusp effect in the decay $K^+\to\pi^0\pi^0\pi^+$
has become one of the most precise methods to extract S-wave pion--pion scattering lengths
from experiment~\cite{Cabibbo,CI,Batley,Gamiz,CGKR,Photons}.  
Very loosely speaking, the cusp in the invariant mass spectrum of the $\pi^0\pi^0$ pair
is generated by the decay  $K^+\to\pi^+\pi^+\pi^-$ followed by charge-exchange
rescattering $\pi^+\pi^-\to\pi^0\pi^0$, plus the fact that the pion mass difference 
shifts the $\pi^+\pi^-$ threshold into the physical region (see also~\cite{MMS}).
What makes this channel particularly apt for an investigation of the cusp,
apart from the enormous statistics collected by the NA48/2 collaboration~\cite{Batley},
is the significantly larger branching fraction of $K^+\to\pi^+\pi^+\pi^-$ compared to
$K^+\to\pi^0\pi^0\pi^+$, such that the perturbation of the decay spectrum of the latter
is very sizeable.  
This is in marked contrast to two other decays that have been studied subsequently and that
display, in principle, cusp structures generated by the same mechanism: 
$K_L \to 3\pi^0$~\cite{CI,Gamiz,BFGKR,KTeV}, and
$\eta \to 3\pi^0$~\cite{BFGKR,Ditsche,Gullstrom,Adolph,Unverzagt,Prakhov}.
In both of these, the weaker coupling to the charged-pion final state diminishes
the cusp to a mere 1--2\% effect on the decay spectrum.

In this respect, $\eta'\to\eta\pi^0\pi^0$ decays offer a more promising candidate for an alternative
channel to study the cusp.\footnote{The cusp in this channel is briefly discussed
in the framework of unitarized chiral perturbation theory in~\cite{nissler}.}  
Indeed, in the isospin limit 
$\textrm{BR}(\eta'\to\eta\pi^+\pi^-) = 2 \,\textrm{BR}(\eta'\to\eta\pi^0\pi^0)$, 
so one expects a sizeable effect on the $\pi^0\pi^0$ decay spectrum.
In this article, we adapt the non-relativistic effective field theory (NREFT) formalism
developed in~\cite{CGKR,BFGKR,Photons} to these channels.
Compared to $K\to3\pi$ and $\eta\to3\pi$ decays, there is a different ``secondary'' final-state 
rescattering channel $\pi\eta$ to take into account. 
In addition, the formalism has to be slightly amended for particles in the final state whose
mass difference ($M_\eta-M_\pi$ in this case) is not small.
This study is very timely with regards to the upcoming high-statistics $\eta'$ experiments 
at ELSA~\cite{Beck}, MAMI-C~\cite{MAMI-C1,MAMI-C2}, WASA-at-COSY~\cite{WasaAdam,WASA}, 
KLOE-at-DA$\Phi$NE~\cite{KLOElett,KLOE}, or BES-III~\cite{BES}, which are expected
to increase the data basis on $\eta'$ decays by orders of magnitude.
\end{sloppypar}

The outline of this article is as follows.  
In Sect.~\ref{sec:EFT}, we present the generalized non-relativistic effective field theory
framework for $\eta'\to\eta\pi\pi$ decays, define the necessary Lagrangians and
perform the matching to Dalitz plot as well as $\pi\pi$ and $\pi\eta$ threshold parameters.
Furthermore, we comment on the effects of inelastic channels.
In Sect.~\ref{sec:Amp}, we present the result for the decay amplitudes up to two loops,
including radiative corrections.
This section comprises the central result of this study and can be used in forthcoming experimental
analyses of these decays.
We turn around the argument and \emph{predict} the size and specific shape of the cusp 
in Sect.~\ref{sec:prediction}.  
Our findings are summarized in Sect.~\ref{sec:summary}.

\section{NREFT for  \boldmath{$\eta'\to\eta\pi\pi$}}\label{sec:EFT}

We consider the neutral and charged decay modes 
\begin{align}
\eta'(P_{\eta'}) &\to \pi^0(p_1)\pi^0(p_2)\eta(p_3) ~, \nn 
\eta'(P_{\eta'}) &\to \pi^+(p_1)\pi^-(p_2)\eta(p_3) ~. 
\end{align}
The charged channel only serves as an ``auxiliary mode'' for the cusp analysis.
The kinematical variables are defined in the usual way $s_i=(P_{\eta'}-p_i)^2$, 
$s_1+s_2+s_3 = \mep + M_1^2+M_2^2+M_3^2$,
with $p_i^2=M_i^2$, $i=1,2,3$. We will use the notation $M_{\pi^{\pm}}=M_{\pi}$ throughout.

We shall proceed along the lines of~\cite{CGKR} to develop the 
modified non-relativistic Lagrangian framework for our calculation. 
It is set up in such a way that the results are manifestly covariant,
with the correct analytic structure of the decay amplitude fully reproduced
in the low-energy region.
A consistent power counting scheme can be constructed in a very similar manner to the one in~\cite{CGKR,BFGKR}:
we introduce the formal non-relativistic parameter $\epsilon$ and count
the pion and $\eta$ 3-momenta (in the $\eta'$ rest frame) as $\Order(\epsilon)$, 
the kinetic energies $T_i=p_i^0-M_i$ as $\Order(\epsilon^2)$, 
and the masses of the particles involved as $\Order(1)$. 
Let us remark, however, that as opposed to the case of $\pi\pi$-scattering, 
where the mass difference between the two scattering particles amounts to a small contribution 
in isospin breaking at maximum, for $\pi\eta$-scattering we count $M_{\eta}-M_{\pi}=\Order(1)$. 
We will thus provide a Lagrangian framework that reproduces the low-energy expansion in a standard manner.
As the momenta of the decay products can be larger than in $K\to 3\pi$ decays,
it is not \emph{a priori} clear how fast this non-relativistic expansion converges 
for $\eta'\to\eta\pi\pi$.  In this sense, the approach is kin to traditional 
Dalitz plot parameterizations in terms of polynomials: one has to include as many terms
as necessary to achieve a good description of the data, which 
we expect to provide with the number of terms given in the following.

\begin{sloppypar}
The loop expansion using this non-relativistic Lagrangian framework produces a correlated
expansion in $\epsilon$ and $\pi\pi$ as well as $\pi\eta$ threshold parameters, 
which we denote summarily by $a_{\pi\pi}$ and $a_{\pi\eta}$, respectively,
or by $a$ in case we refer generically to both.
Each two-particle-rescattering increases the order of the loop contribution
by $a\epsilon$~\cite{CGKR}.
\end{sloppypar}

\subsection{Non-relativistic Lagrangians}

The full non-relativistic Lagrangian can be split up into separate parts describing the 
$\eta'\to \eta \pi \pi$ tree amplitude and the $\pi\pi$ and $\pi\eta$ final-state interactions:
\beq
\Lagr=\Lagr_{\eta'}+\Lagr_{\pi\pi}+\Lagr_{\pi\eta}~.
\eeq
The $\eta'\to\eta\pi\pi$ Lagrangian up to $\Order(\epsilon^4)$ is given by
\begin{align}
\Lagr_{\eta'} &=\ 2\eta'^\dagger W_{\eta'}\left(i\partial_t-W_{\eta'}\right)\eta' \nn
&+ \frac{1}{2}\sum_{i=0}^2 G_i\left(\eta'^{\dagger}\left(W_{\eta}-M_{\eta}\right)^i\eta \Phi_0\Phi_0 + h.c. \right)\nn
&+G_3\left(\eta'^\dagger\eta\left(W_0^2\Phi_0\Phi_0-W_0\Phi_0W_0\Phi_0\right)+h.c.\right) \nn
&+\sum_{i=0}^2 H_i\left(\eta'^{\dagger}\left(W_{\eta}-M_{\eta}\right)^i\eta \Phi_+\Phi_- + h.c. \right)\nn
&+H_3\Bigl(\eta'^\dagger \eta \bigl(W_{\pm}^2\Phi_+\Phi_-+\Phi_+W_{\pm}^2\Phi_- \nn
& \qquad -2W_{\pm}\Phi_+W_{\pm}\Phi_-\bigr)+h.c.\Bigr) +\ldots ~,
\end{align}
where $\Phi$, $\eta$, $\eta'$ denote non-relativistic pion, $\eta$, and $\eta'$ field operators,
$W_{a}=\sqrt{M_{a}^2-\Delta}$, where $\Delta$ is the Laplacian, 
and $G_i$, $H_i$ are the low-energy couplings in the neutral and charged channel, respectively. 
The ellipsis stands for higher orders in the $\epsilon$ expansion. 

We start with $\pi\pi$ final-state interactions and consider the three channels in $(i)$ 
$(\pi^a\pi^b\to\pi^c\pi^d)$: (00) $(00;00)$, $(x)$ $(+-;00)$, $(+-)$ $(+-;+-)$. 
For the following discussion we introduce the notation
\begin{align}
(\Phi_n)_{\mu}         &=(\p_n)_\mu\Phi_n ~,                & 
(\Phi_n)_{\mu\nu}      &=(\p_n)_{\mu}(\p_n)_{\nu}\Phi_n ~,\nn
(\Phi_n^{\dagger})_{\mu} &=(\p_n^{\dagger})_{\mu}\Phi_n^{\dagger}~,&
(\Phi_n^{\dagger})_{\mu\nu}&=(\p_n^{\dagger})_{\mu}(\p_n^{\dagger})_{\nu}\Phi_n^{\dagger}~,\nn
(\p_n)_{\mu} & =(W_n,-i\nabla) ~, & (\p_n^{\dagger})_{\mu}&=(W_n,+i\nabla) ~,
\end{align}
for $n=a,b,c,d$. Analogous definitions hold for derivatives on the $\eta$ field, with $\Phi\to\eta$ and
$\p_n\to\p_\eta$.
The $\pi\pi$ final-state Lagrangian can be written as~\cite{BFGKR}
\begin{align}
\Lagr_{\pi\pi}=2\underset{k=0,\pm}{\sum}\Phi_k^{\dagger}W_k\left(i\partial_t-W_k\right)\Phi_k+\underset{i}{\sum}\Lagr_i~,
\end{align}
where the first part is the free pion propagator and
\begin{align}
\mathcal{L}_i&=x_iC_i\left(\Phi_c^{\dagger}\Phi_d^{\dagger}\Phi_a\Phi_b+h.c.\right)\nn
&+x_iD_i\Big\{(\Phi_c^{\dagger})_{\mu}(\Phi_d^{\dagger})^{\mu}\Phi_a\Phi_b+\Phi_c^{\dagger}\Phi_d^{\dagger}(\Phi_a)_{\mu}(\Phi_b)^{\mu}\nn
&\qquad-h_i\Phi_c^{\dagger}\Phi_d^{\dagger}\Phi_a\Phi_b+h.c.\Big\}\nn
&+x_iF_i\Big\{(\Phi_c^{\dagger})_{\mu\nu}(\Phi_d^{\dagger})^{\mu\nu}\Phi_a\Phi_b+\Phi_c^{\dagger}\Phi_d^{\dagger}(\Phi_a)_{\mu\nu}(\Phi_b)^{\mu\nu}\nn
&\qquad+2(\Phi_c^{\dagger})_{\mu}(\Phi_d^{\dagger})^{\mu}(\Phi_a)_{\nu}(\Phi_b)^{\nu}+h_i^2\Phi_c^{\dagger}\Phi_d^{\dagger}\Phi_a\Phi_b\nn
&\qquad-2h_i\left((\Phi_c^{\dagger})_{\mu}(\Phi_d^{\dagger})^{\mu}\Phi_a\Phi_b+\Phi_c^{\dagger}\Phi_d^{\dagger}(\Phi_a)_{\mu}(\Phi_b)^{\mu}\right)\nn
&\qquad+h.c.\Big\}+\ldots~,
\end{align}
with $x_{00}=1/4$, $x_{x}=x_{+-}=1$ and $h_i=s_i^t-\frac{1}{2}(M_a^2+M_b^2+M_c^2+M_d^2)$, 
where $s_i^t$ is the physical threshold of the {\it i}th channel, 
explicitly $h_{00}=2M_{\pi^0}^2$, $h_{x}=3M_{\pi}^2-M_{\pi^0}^2$, $h_{+-}=2M_{\pi}^2$. 
We have omitted P-wave contributions (in the $(+-)$ channel) as they do not contribute in 
$\eta'\to\eta\pi\pi$ as long as conservation of C-parity is assumed.
The ellipsis denotes the omission of higher-order terms in the $\epsilon$ expansion.

In the case of $\pi\eta$ scattering, we consider the channels
$(i)$ $(\eta\pi^a\to\eta\pi^a)$: $(\eta0)$ $(\eta0;\eta0)$, $(\eta+)$ $(\eta+;\eta+)$ 
(the $\eta\pi^-\to\eta\pi^-$ amplitude is identical to $\eta\pi^+\to\eta\pi^+$ by charge conjugation).
For the $\pi\eta$ Lagrangian we find
\begin{equation}
\Lagr_{\pi\eta}=2\eta^\dagger W_{\eta}\left(i\partial_t-W_{\eta}\right)\eta+\sum_i\Lagr_{i}~,
\end{equation}
where the first term is again the free particle propagator. 
Before giving the explicit form of the interaction piece, we define a differential operator
\begin{align}
\hat{s}^{-1} \doteq \Bigl[\me+M_a^2+(\p_\eta^\dagger)_{\mu}(\p_a^\dagger)^{\mu}+(\p_\eta)_{\mu}(\p_a)^{\mu}\Bigr]^{-1}.
\end{align}
This operator has to be understood as follows: 
expand about the respective thresholds $s_{\eta\pi^0}^t=(M_{\eta}+M_{\pi^0})^2$ and $s_{\eta \pi}^t=(M_{\eta}+M_{\pi})^2$,
\begin{align}\nonumber
&\hat{s}^{-1}\eta^{\dagger}\Phi_a^{\dagger}\eta\Phi_a+h.c.=\frac{1}{(M_{\eta}+M_a)^2}\left(\eta^{\dagger}\Phi_a^{\dagger}\eta\Phi_a+h.c.\right)\\\nonumber
&+\frac{1}{(M_{\eta}+M_a)^4}\Bigl((\eta^{\dagger})_{\mu}(\Phi_a^{\dagger})^{\mu}\eta\Phi_a+\eta^{\dagger}\Phi_a^{\dagger}(\eta)_{\mu}(\Phi_a)^{\mu}\\\nonumber
&\qquad-s_{\eta a}^t\eta^{\dagger}\Phi_a^{\dagger}\eta\Phi_a+h.c.\Bigr)+\ldots~,
\end{align}
then apply the Feynman rules of the theory in momentum space and resum the result. 
Note that the differential operator $\hat{s}^{-1}$ does not violate analyticity and unitarity of the S-Matrix
in the low-energy region, since it is obeyed term by term.

We can now display the interaction piece,
\begin{align}
\Lagr_i=\ 
 &C_i\left(\Phi_a^{\dagger}\eta^{\dagger}\Phi_a\eta+h.c.\right)\nn 
+\ &D_i\Big\{(\eta^{\dagger})_{\mu}(\Phi_a^{\dagger})^{\mu}\eta\Phi_a+\eta^{\dagger}\Phi_a^{\dagger}(\eta)_{\mu}(\Phi_a)^{\mu}-h_i\eta^{\dagger}\Phi_a^{\dagger}\eta\Phi_a \nn
&\qquad+\Delta_{\eta a}^2\hat{s}^{-1}\eta^{\dagger}\Phi_a^{\dagger}\eta\Phi_a+h.c.\Big\} \nn
+\ &\frac{E_i}{2}\Big\{\Big(\eta^{\dagger}(\Phi_a^{\dagger})_\mu-(\eta^{\dagger})_\mu\Phi_a^{\dagger}\Big)
\Big((\eta)^\mu\Phi_a+\eta (\Phi_a)^\mu\Big)\nn
&\qquad-\Delta_{\eta a}^2\hat{s}^{-1}\eta^{\dagger}\Phi_a^{\dagger}\eta\Phi_a+h.c.\Big\}+\ldots~,
\end{align}
with $h_{\eta 0}=M_{\eta}^2+M_{\pi^0}^2$, $h_{\eta +}=M_{\eta}^2+M_{\pi}^2$,
and $\Delta_{\eta a} = \me - M_a^2$. The ellipsis again stands for higher-order terms in $\epsilon$. 
We do not consider six-particle couplings, since their contribution to the amplitude is negligible 
(see Sect.~\ref{sec:6part}).

\subsection{Matching}

\begin{sloppypar}
We obtain the couplings of $\Lagr_{\eta'}$ by matching to the standard Dalitz plot distribution
\beq
|\M(x,y)|^2 = |\N|^2 \bigl(1+ay+by^2+dx^2 + \ldots \bigr) ~,\label{eq:DalitzDef}
\eeq
where $\N$ is a normalization constant, and
\begin{align}
x&=\frac{\sqrt{3}(s_1-s_2)}{2M_{\eta'}Q_{\eta'}} = \frac{\sqrt{3}(p_2^0-p_1^0)}{Q_{\eta'}} ~,\nn
y&=\frac{\left(M_{\eta}+2M_{\pi^0}\right)[\left(M_{\eta'}-M_{\eta}\right)^2-s_3]}{2M_{\eta'}M_{\pi^0}Q_{\eta'}}-1 \nn
 &=\frac{M_{\eta}+2M_{\pi^0}}{M_{\pi^0}Q_{\eta'}}\bigl(p_3^0-M_\eta\bigr)-1 ~,\nn
Q_{\eta'}&=M_{\eta'}-M_{\eta}-2M_{\pi^0} ~,\label{eq:Dalitz}
\end{align}
and we have omitted a C-violating term $\propto x$.
The particle energies $p_i^0$ in the $\eta'$ rest frame are related to the invariants $s_i$ according to
\beq
p_i^0 = \frac{\mep+M_i^2-s_i}{2M_{\eta'}} ~.
\eeq
For the charged channel we have to replace $M_{\pi^0} \to M_\pi$ in~\eqref{eq:Dalitz}.
Equation~\eqref{eq:DalitzDef} is reproduced, up to higher orders in $x$, $y$, by the following polynomial amplitude:
\beq
\M(x,y) = \N \biggl\{ 1+\frac{a}{2}y +\frac{1}{2}\Bigl(b-\frac{a^2}{4}\Bigr)y^2+ \frac{d}{2}x^2 + \ldots \biggr\} ~,
\eeq
from which one can obtain the 
matching to the coupling constants $G_i$ according to
\begin{align}
G_0&=\N\biggl\{1-\frac{a}{2}+\frac{1}{2}\Bigl(b-\frac{a^2}{4}\Bigr)\biggr\}~,\nn
G_1&=\N\biggl\{\frac{a}{2}-\Bigl(b-\frac{a^2}{4}\Bigr)\biggr\}~\frac{M_\eta+2M_{\pi^0}}{M_{\pi^0}Q_{\eta'}}~, \label{NCTreematching}\\
G_2&=\N\Bigl(b-\frac{a^2}{4}\Bigr)\frac{(M_\eta+2M_{\pi^0})^2}{2M_{\pi^0}^2Q_{\eta'}^2}~,\quad
G_3=\N\frac{3d}{2Q_{\eta'}^2}~.\nonumber
\end{align}
In the isospin limit, the charged-channel couplings $H_i$ are related to the $G_i$ via
$H_i=-\sqrt{2}G_i$ (in the Condon--Shortley phase convention), which we will assume in the following.

The couplings $C_i$, $D_i$, $E_i$, $F_i$ are obtained by matching to the effective range expansion
of $\pi\pi$ and $\pi\eta$ scattering.
The partial wave decomposition of the $\pi\pi$ scattering amplitude is 
conventionally written as~\cite{ACGL}
\beq\label{partamp}
T^I(s,t)=32\pi\sum_{l=0}^\infty(2l+1)t_l^I(s)P_l(z)~,
\eeq
where $t_l^I(s)$ is the partial wave amplitude of angular momentum $l$ and isospin $I$, 
$P_l(z)$ are the Legendre polynomials, and $z=\cos\theta$ is the scattering angle in the center-of-mass system. 
Close to threshold of the pertinent channel, one can perform an expansion in the center-of-mass momentum 
$q^2_{ab}(s)=\lambda(s,M_a^2,M_b^2)/4s$ 
(with the standard K\"all\'en function $\lambda(a,b,c)=a^2+b^2+c^2-2(a b+a c+b c)$)
according to
\beq\label{partampexp}
\text{Re }t_l^I(s)=q_{ab}^{2l}\Bigr\{a_l^I+b_l^Iq_{ab}^2+f_l^Iq_{ab}^4+\Order(q_{ab}^6)\Bigr\}~.
\eeq
As only $\pi\pi$ S-waves are considered in the following, we will use the slightly simplified
notation $a_0$, $a_2$ for the S-wave scattering lengths of isospin 0 and 2, and similarly 
for the effective ranges $b_i$ and shape parameters $f_i$.

We use a definition of the $\pi\eta$ threshold parameters strictly analogous 
to~\eqref{partamp}, \eqref{partampexp}, and denote the S- and P-wave scattering lengths and
the S-wave effective range by $\bar a_0$, $\bar a_1$, and $\bar b_0$.
These quantities are modified compared to the more conventional parameterization used in 
$\pi\eta$ scattering~\cite{BKM:pieta}; our $\pi\eta$ threshold parameters are related to the 
conventional ones $a_0^{\pi\eta}$, $a_1^{\pi\eta}$, $b_0^{\pi\eta}$ by
\beq
\bar{a}_i=\frac{M_{\eta}+M_{\pi}}{4}a_i^{\pi\eta} \,, ~
\bar{b}_0=\frac{M_{\eta}+M_{\pi}}{4}\left(b_0^{\pi\eta}+\frac{a_0^{\pi\eta}}{2M_{\pi}M_{\eta}}\right) \,.
\eeq
We refrain from including the shape parameter of $\pi\eta$-scattering: 
its contribution is expected to be tiny, 
and the threshold parameters of $\pi\eta$-scattering are systematically smaller than those of $\pi\pi$-scattering; 
c.f.\ Appendix~\ref{pietathresh}.
Furthermore, even the leading threshold parameters of $\pi\eta$-scattering are not easy to come by 
within a sensible error range; see~\cite{Kolesar} and Appendix~\ref{pietathresh}.
However, the inclusion of higher orders in the modified non-relativistic framework is straightforward 
and can be easily performed, should the threshold parameters be obtained to greater accuracy.
\end{sloppypar}

We can fix the couplings of the non-relativistic Lagrangian by matching the amplitude of the effective theory 
to~\eqref{partampexp}. The $\pi\pi$ and $\pi\eta$ Lagrangians generate the following tree amplitudes:
\begin{align}
\text{Re }T^{\pi\pi}_i&=2C_{i}+8D_{i}q_{ab}^2+32F_{i}q_{ab}^4+\ldots~,\\
\text{Re }T^{\eta a}&=2C_{\eta a}+8D_{\eta a}q_{\eta a}^2+E_{\eta a}\bigg(t-u+\frac{\Delta_{\eta a}^2}{s}\bigg)+\ldots~.\nonumber
\end{align}
The matching conditions can be simply read off: for $\pi\pi$ scattering we have~\cite{CGKR,BFGKR}
\begin{align}
2C_{00}&=\frac{N}{3}(a_0+2a_2)(1-\xi)~, & 
8D_{00}&=\frac{N}{3}(b_0+2b_2)~,\nn
2C_{x} &=\frac{N}{3}(a_2-a_0)\Big(1+\frac{\xi}{3}\Big)~, &
8D_{x} &=\frac{N}{3}(b_2-b_0)~,\nn
2C_{+-}&=\frac{N}{6}(2a_0+a_2)(1+\xi)~, &
8D_{+-}&=\frac{N}{6}(2b_0+b_2)~,\label{eq:pipimatch}
\end{align}
and the matching conditions for the $F_i$ are identical to those for the $D_i$ with
the replacements $b_i \to f_i/4$.
Isospin breaking in the S-wave scattering lengths has been taken into account 
at leading order in chiral perturbation theory~\cite{KnechtUrech} in~\eqref{eq:pipimatch}, 
$\xi=(\mpi-\mpn)/\mpi$, and $N=32\pi$.
For $\pi\eta$ scattering we find
\begin{align}
2C_{\eta 0} &= 2C_{\eta +} = N\bar{a}_0~, &
8D_{\eta 0} = 8D_{\eta +} = N\bar{b}_0~, \nn
4E_{\eta 0} &= 4E_{\eta +} =3N\bar{a}_1~. &
\end{align}

\subsection{Six-particle vertices, inelastic channels}\label{sec:6part}
\begin{figure}
 \centering
 \includegraphics[width=0.7\linewidth]{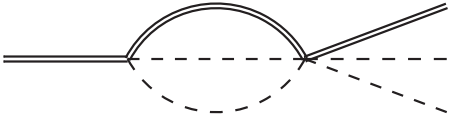}
 \caption{Non-relativistic two-loop graph involving the six-particle vertex.
Double lines denote $\eta'$ and $\eta$, the dashed line the $\pi^0$.
 \label{fig:six}}
\end{figure}
Six-particle interactions can be included in the modified non-relativistic framework by straightforward extension
\beq
\Lagr_{\eta\pi\pi}=\frac{1}{4}F_0\eta^\dagger(\Phi_0^\dagger)^2\eta\Phi_0^2
+F_0'\eta^\dagger\Phi_+^\dagger\Phi_-^\dagger\eta\Phi_+ \Phi_- + \ldots~,
\eeq
where the ellipsis denotes terms with derivative couplings. 
To give a rough estimate of the contribution to the neutral channel, we perform a threshold expansion for the diagram 
in Fig.~\ref{fig:six}, which amounts to the application of the ``classical'' 
non-relativistic framework~\cite{HadAtom}. 
The real part is a constant (albeit divergent), which can be absorbed in a redefinition of the coupling $G_0$ 
and thus simply amounts to a change of the renormalization prescription. 
The imaginary part of the diagram is given by
\beq
\text{Im}\mathcal{A}_N^{\pi^0\pi^0\eta} = \frac{F_0G_0}{256\pi^2}
\frac{M_{\pi^0}M_\eta^{1/2}}{(M_{\eta}+2M_{\pi^0})^{3/2}}Q_{\eta'}^2 + \Order(Q_{\eta'}^3).
\eeq
We can give an estimate for $F_0$ by matching to chiral perturbation theory. 
At lowest order (assuming isospin symmetry), we find
\beq
F_0 = \frac{M_\pi^3}{36\Fpi^4}\biggl\{\frac{5M_\pi}{\me}-\frac{1}{M_\pi+M_\eta} \biggr\}
~,
\eeq
where $F_\pi$ is the pion decay constant. 
The imaginary part of the diagram can be mimicked
by allowing $G_0$ to have a small imaginary part.  Numerically,
\beq
\frac{\text{Im}\,G_0}{\text{Re}\,G_0} \simeq 10^{-6},
\eeq
which is sufficiently small to assume the coupling constants of the non-relativistic Lagrangian to be real.

\begin{figure}
 \centering
 \includegraphics[width=0.9\linewidth]{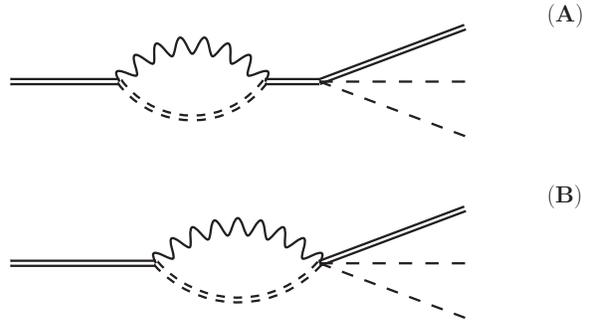}
 \caption{Contributions of the inelastic $\rho^0\gamma$ channel.
The double-dashed line denotes the $\rho^0$, the wiggly line a photon; 
otherwise, the line style is as in Fig.~\ref{fig:six}.}
 \label{fig:inelastic}
\end{figure}
One may also wonder about effects of the finite width of the $\eta'$, or the coupling
to inelastic channels other than $\eta\pi\pi$.  
One effect of the finite width is that the mass of the $\eta'$ obtains an imaginary part;
however, in view of $\Gamma_{\eta'}/M_{\eta'} \approx 2\times 10^{-4}$, this is still negligible
if we aim at an overall accuracy at the percent level at best.
The largest decay channel of the $\eta'$ other than $\eta\pi\pi$ is $\rho^0\gamma$~\cite{pdg},
which could contribute via the diagrams shown in Fig.~\ref{fig:inelastic}.
The rescattering vertex in diagram (B) is anomalous, but there is no
$VV\to\eta\pi^0\pi^0$ vertex in the Wess--Zumino--Witten Lagrangian~\cite{WZ71,Wi83}, so 
we neglect this term.  Diagram (A) leads to a complex wave-function renormalization factor $Z$,
for which we find
\beq
\textrm{Im}\,Z_{\rho^0\gamma} = \frac{2\mep+M_\rho^2}{\mep-M_\rho^2} \frac{\Gamma_{\eta'\to\rho^0\gamma}}{M_{\eta'}}
\approx 5\times 10^{-4} ~.
\eeq
This is therefore also an extremely small correction, which can furthermore be absorbed in an overall phase. 
We conclude that, for the purpose of this investigation, we can assume real coupling constants
and neglect inelastic channels.

\section{The decay amplitudes to two loops}\label{sec:Amp}

We use the following decomposition of the amplitudes:
\begin{align}
\M_{\eta'\to\eta\pi^0\pi^0}=\M_N^\textrm{tree}+\M_N^\textrm{1-loop}+\M_N^\textrm{2-loop}+\ldots,\nn
\M_{\eta'\to\eta\pi^+\pi^-}=\M_C^\textrm{tree}+\M_C^\textrm{1-loop}+\M_C^\textrm{2-loop}+\ldots,
\end{align}
to underline that the tree amplitude of $\mathcal{L}_{\eta'\to\eta\pi\pi}$ is modified by final-state interactions of one, two, etc. loops.

\subsection{Tree amplitudes}

The tree amplitudes are given by
\begin{align}\label{NCTree}\nonumber
\Amp_N^\textrm{tree}(s_1,s_2,s_3)&= \sum_{i=0}^2 G_iX_3^i + G_3(X_1\!-\!X_2)^2 ,\\
\Amp_C^\textrm{tree}(s_1,s_2,s_3)&= \sum_{i=0}^2 H_iX_3^i + H_3(X_1\!-\!X_2)^2 ,
\end{align}
where $X_k=p_k^0-M_\eta$, $k=1,\,2,\,3$.

\subsection{One-loop amplitudes}

For the one-loop amplitudes we find
\begin{align}
\M_N^\textrm{1-loop}&(s_1,s_2,s_3)=\mathcal{B}_{N1}(s_3)J_{+-}(s_3)+\mathcal{B}_{N2}(s_3)J_{00}(s_3)\nn
&+\big\{\mathcal{B}_{N3}(s_1,s_2,s_3)J_{\eta 0}(s_1)+(s_1\leftrightarrow s_2)\big\}~, \nn
\M_C^\textrm{1-loop}&(s_1,s_2,s_3)=\mathcal{B}_{C1}(s_3)J_{+-}(s_3)+\mathcal{B}_{C2}(s_3)J_{00}(s_3)\nn 
&+\big\{\mathcal{B}_{C3}(s_1,s_2,s_3)J_{\eta +}(s_1)+(s_1\leftrightarrow s_2)\big\}~,
\end{align}
with the one-loop function
\beq
J_{ab}(s_k)=\frac{iq_{ab}(s_k)}{8\pi\sqrt{s_k}} ~, 
\eeq
and the polynomials
\begin{align}
\mathcal{B}_{N1}&(s_3)= 2C_x(s_3)\bigg\{ \sum_{i=0}^2 H_i X_3^i 
+ H_3\frac{4{\bf Q}^2_3}{3s_3}q_{+-}^2(s_3)\bigg\} ~, \nn
\mathcal{B}_{N2}&(s_3) = C_{00}(s_3)\bigg\{ \sum_{i=0}^2 G_i X_3^i
+ G_3\frac{4{\bf Q}^2_3}{3s_3}q_{00}^2(s_3)\bigg\} ~, \nn
\mathcal{B}_{N3}&(s_1,s_2,s_3) = 2C_{\eta 0}(s_1)\bigg\{G_0+G_1Z_{1,+}^{\eta 0} \nn
&+G_2\biggl[\bigl(Z_{1,+}^{\eta 0}\bigr)^2+\frac{{\bf Q}^2_1}{3s_1}q_{\eta 0}^2(s_1)\biggr] \nn
&+G_3\biggr[\bigl(Z_{1,-}^{\eta 0}-X_1\bigr)^2+\frac{{\bf Q}^2_1}{3s_1}q_{\eta 0}^2(s_1)\biggr]\bigg\} \nn
&-E_{\eta 0}\frac{q_{\eta 0}^2(s_1)}{3M_{\eta'}}\biggl[s_3-s_2+\frac{\Delta_{\eta 0}}{s_1}\bigl(M_{\eta'}^2-M_{\pi^0}^2\bigr)\biggr] \nn
&\quad\times\Big\{G_1+2G_2Z_{1,+}^{\eta 0}+2G_3\bigl(X_1-Z_{1,-}^{\eta 0}\bigr)\Big\} 
\end{align}
for the neutral channel, and 
{\allowdisplaybreaks 
\begin{align} 
\mathcal{B}_{C1}&(s_3)= 2C_{+-}(s_3)\bigg\{ \sum_{i=0}^2 H_i X_3^i
+ H_3\frac{4{\bf Q}^2_3}{3s_3}q_{+-}^2(s_3)\bigg\} ~,\nn
\mathcal{B}_{C2}&(s_3) = C_{x}(s_3)\bigg\{ \sum_{i=0}^2 G_i X_3^i 
+G_3\frac{4{\bf Q}^2_3}{3s_3}q_{00}^2(s_3)\bigg\} ~,\nn
\mathcal{B}_{C3}&(s_1,s_2,s_3) = 2C_{\eta +}(s_1)\bigg\{H_0+H_1Z_{1,+}^{\eta +}\nn
&+H_2\biggl[\bigl(Z_{1,+}^{\eta +}\bigr)^2+\frac{{\bf Q}^2_1}{3s_1}q_{\eta +}^2(s_1)\biggr]\nn
&+H_3\biggr[\big(Z_{1,-}^{\eta +}-X_1\big)^2+\frac{{\bf Q}^2_1}{3s_1}q_{\eta +}^2(s_1)\biggr]\bigg\}\nn
&-E_{\eta +}\frac{q_{\eta +}^2(s_1)}{3M_{\eta'}}\bigg[s_3-s_2+\frac{\Delta_{\eta +}}{s_1}\big(M_{\eta'}^2-M_{\pi}^2\big)\bigg]\nn
&\quad\times\Big\{H_1+2H_2Z_{1,+}^{\eta +}+2H_3\big(X_1-Z_{1,-}^{\eta +}\big)\Big\} 
\end{align}}\noindent
for the charged channel. The following abbreviations have been used:
\begin{align}
Q_a^0&=p_b^0+p_c^0~(\text{+cycl.})~, \quad{\bf Q}_a^2=\frac{\lambda(M_{\eta'}^2,M_a^2,s_a)}{4M_{\eta'}^2}~,\nn
Z_{k,\pm}^{ab}&=\frac{Q_k^0}{2}\left(1\pm\frac{\Delta_{ab}}{s_k}\right)-M_{a}~,\\
C_{bc}(s_a) &= C_{bc}+4D_{bc}q_{bc}^2(s_a)+16F_{bc}q_{bc}^4(s_a) \,,~ q_x=q_{+-} ~. \nonumber
\end{align}
Note in particular that there are no higher orders in the $\pi\eta$ mass difference omitted 
in the P-wave of $\pi\eta$ scattering;
with our Lagrangian definition of the P-wave operators, the formulae above are exact.

\subsection{Two-loop amplitudes}

\begin{figure}
 \centering
 \includegraphics[width=0.9\linewidth]{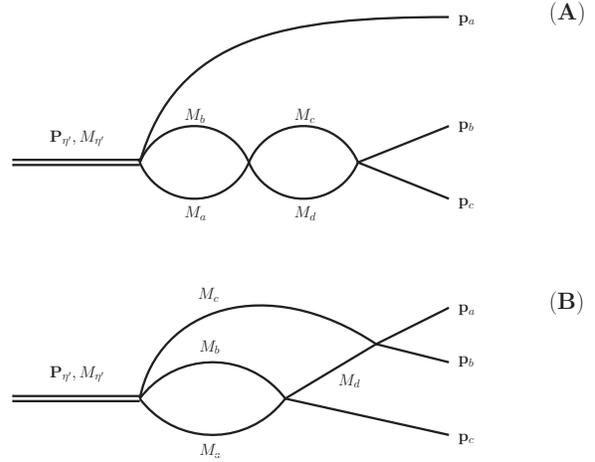}
 \caption{The two distinct topologies appearing at two loops.
  The first one (A) is a simple product of one-loop functions, 
  the second one (B) yields a more complicated analytic behavior.
  Here, the double line denotes the $\eta'$, while single lines stand generically
  for any of the particles in the final state $\eta$, $\pi^\pm$, and $\pi^0$.}
 \label{fig:twoloopbig}
\end{figure}
\begin{sloppypar}
The two-loop amplitudes contain diagrams of two distinct topologies; see Fig.~\ref{fig:twoloopbig}. 
We find the following: 
{\allowdisplaybreaks
\begin{align}
\M&_N^\textrm{2-loop}(s_1,s_2,s_3)=C_{00}(s_3)\mathcal{B}_{N2}(s_3)J_{00}^2(s_3)\nn
&+\Bigl[C_{00}(s_3)\mathcal{B}_{N1}(s_3)+2C_{x}(s_3)\mathcal{B}_{C2}(s_3)\Bigr]J_{00}(s_3)J_{+-}(s_3)\nn
&+2C_{x}(s_3)\mathcal{B}_{C1}(s_3)J_{+-}^2(s_3)\nn
&+\Bigl\{2C_{\eta 0}(s_1)\mathcal{B}_{N3}(s_1,s_2,s_3)J_{\eta 0}^2(s_1) \nn
&\quad+2G_0C_{00}C_{\eta 0}F_0(M_{\pi^0},M_{\pi^0},M_{\eta},M_{\pi^0},s_1)\nn
&\quad+4H_0C_{x}C_{\eta 0}F_0(M_{\pi},M_{\pi},M_{\eta},M_{\pi^0},s_1)\nn
&\quad+4G_0C_{\eta 0}^2F_0(M_{\eta},M_{\pi^0},M_{\pi^0},M_{\eta},s_1)+(s_1\leftrightarrow s_2)\Bigr\}\nn
&+4G_0C_{\eta 0}C_{00}F_{\eta}(M_{\eta},M_{\pi^0},M_{\pi^0},M_{\pi^0},s_3) \nn
&+8H_0C_{\eta +}C_{x}F_{\eta}(M_{\eta},M_{\pi},M_{\pi},M_{\pi},s_3)  \label{neutraltwoloop}
\end{align}}\noindent
in the neutral channel, and
\begin{align}
\M&_C^\textrm{2-loop}(s_1,s_2,s_3)=2C_{+-}(s_3)\mathcal{B}_{C1}(s_3)J_{+-}^2(s_3)\nn
&+\!\Bigl[C_{x}(s_3)\mathcal{B}_{N1}(s_3) \!+\! 2C_{+-}(s_3)\mathcal{B}_{C2}(s_3)\Bigr]J_{00}(s_3)J_{+-}(s_3)\nn
&+C_{x}(s_3)\mathcal{B}_{N2}(s_3)J_{00}^2(s_3)\nn
&+\Bigl\{2C_{\eta +}(s_1)\mathcal{B}_{C3}(s_1,s_2,s_3)J_{\eta +}^2(s_1) \nn
&\quad+4H_0C_{+-}C_{\eta +}F_+(M_{\pi},M_{\pi},M_{\eta},M_{\pi},s_1)\nn
&\quad+2G_0C_{x}C_{\eta +}F_+(M_{\pi^0},M_{\pi^0},M_{\eta},M_{\pi},s_1)\nn
&\quad+4H_0C_{\eta +}^2F_+(M_{\eta},M_{\pi},M_{\pi},M_{\eta},s_1)+(s_1\leftrightarrow s_2)\Bigr\}\nn
&+8H_0C_{\eta +}C_{+-}F_{\eta}(M_{\eta},M_{\pi},M_{\pi},M_{\pi},s_3)\nn
&+4G_0C_{\eta 0}C_{x}F_{\eta}(M_{\eta},M_{\pi^0},M_{\pi^0},M_{\pi^0},s_3) \label{chargedtwoloop}
\end{align}
in the charged channel. 
The analytic form of the genuine two-loop function $F_k(M_a,M_b,M_c,M_d,s_k)$ can be found in Appendix~\ref{twoloopfunc}.
The representation above is valid to 
$\Order(a_{\pi\pi}^2\epsilon^6,a_{\pi\pi}a_{\pi\eta}\epsilon^2,a_{\pi\eta}^2\epsilon^2)$.
Due to the smallness of the $\pi\eta$ rescattering effects, this is well justified.
\end{sloppypar}

\subsection{Radiative corrections}\label{sec:rad}

\begin{figure}
 \centering
 \includegraphics[width=0.7\linewidth]{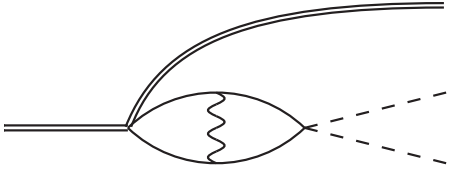}
 \caption{Leading contribution to radiative corrections in $\eta'\to\eta\pi^0\pi^0$.
The wiggly line denotes the exchange of a Coulomb photon between two charged pions (single full lines).}
 \label{fig:rad}
\end{figure}
\begin{sloppypar}
In~\cite{Photons}, radiative corrections to $K\to3\pi$ decays have been discussed
in the framework of non-relativistic effective field theory.  
The results obtained there for the decay channels $K_L \to 3\pi^0$ and
$K_L \to \pi^+\pi^-\pi^0$ can be adapted immediately to $\eta'\to\eta\pi\pi$ decays, 
so we only quote the final formulae and refer to~\cite{Photons} for the derivation.

In the neutral decay $\eta'\to\eta\pi^0\pi^0$, no ``external'' radiative corrections exist. 
The leading electromagnetic contributions (of $\Order(a_{\pi\pi}\log\epsilon)$) are due to 
virtual-photon exchange inside a charged-pion loop; see Fig.~\ref{fig:rad}.
These become important very close to threshold, as they modify the analytic structure
near the cusp by adding a logarithmic singularity to the square-root-like behavior.
The diagram in Fig.~\ref{fig:rad} can be taken into account by the following replacement
of the charged-pion one-loop function:
\begin{align}\label{eq:replacementCoulomb}
 J_{+-}(s_3) &\to J_{+-}(s_3)+\bar J_C(s_3) ~,\nn
\bar J_C(s_3) &= -\frac{\alpha}{32\pi}\log\bigg(-\frac{4q_{+-}^2(s_3)}{M_\pi^2}\bigg) ~.
\end{align}

As $\eta'\to\eta\pi^+\pi^-$ only serves as the ``auxiliary channel'' in the cusp analysis, 
only radiative corrections of $\Order(e^2a^0)$ are considered, hence we neglect photon exchange inside loops.
The external photon corrections, comprising virtual-photon exchange 
as well as real-photon radiation up to a maximal photon energy $E^*$, can be subsumed in 
the correction factor $\Omega_C(s_3,E^*)$ that multiplies the decay spectrum,
which is given in the soft-photon approximation by
\begin{align}
\frac{d\Gamma}{ds_3} \biggr|_{E_\gamma<E^*} &= 
\Omega_C(s_3,E^*) \frac{d\Gamma^{\rm int}}{ds_3} ~, \label{eq:Omega+-0}\\
\Omega_C(s_3,E^*) &= 1+\frac{\alpha}{\pi} 
\biggl\{ \frac{\pi^2(1+\sigma^2)}{2\sigma} 
+ \frac{8}{3}\sigma^2 \Bigl[ \log\frac{2E^*}{M_\pi} -\frac{1}{3}\Bigr] 
\biggr\} \,,\nonumber
\end{align}
where $\sigma=\sqrt{1-4\mpi/s_3}$.
To the accuracy considered here, ${d\Gamma^{\rm int}}/{ds_3}$ is the $\eta'\to\eta\pi^+\pi^-$
decay spectrum without photon corrections.
See~\cite{Photons} for the more elaborate result without the soft-photon approximation.
\end{sloppypar}

\section{Prediction of the cusp}\label{sec:prediction}

In the previous section, we have given the representation of the $\eta'\to\eta\pi\pi$ decay
amplitudes to two-loop order in terms of a set of coupling constants $G_i$, $H_i$, 
as well as $\pi\pi$ and $\pi\eta$ threshold parameters.
In order to extract the $\pi\pi$ scattering length combination $a_0-a_2$, one would 
have to fit the latter together with the couplings $G_i$ (and possibly even $H_i$ if one wishes
to relax the assumption on isospin conservation in the polynomial terms) to experimental data.
Here we wish to \emph{predict} the cusp in $\eta'\to\eta\pi^0\pi^0$.  For this purpose, 
we first have to fix parameters.

\begin{sloppypar}
The $\pi\pi$ scattering lengths and effective ranges are set to the theoretical values
$a_0=0.220 \pm 0.005$, $a_2 = -0.0444\pm 0.0010$, $b_0 = (0.276 \pm 0.006)\times M_\pi^{-2}$, 
$b_2= (-0.0803 \pm 0.0012)\times M_\pi^{-2}$~\cite{CGL}, and the shape parameters $f_0$, $f_2$
to 0.  
There is no experimental information on the $\pi\eta$ threshold parameters, and,
as discussed in Appendix~\ref{pietathresh}, theoretical constraints from chiral perturbation
theory are not very restrictive.  We will therefore discuss the uncertainties induced by
this lack of knowledge by varying these parameters in the ranges 
$\bar a_0 = (0 \ldots +16)\times 10^{-3}$, 
$\bar b_0 = (0 \ldots +10) \times 10^{-3} M_\pi^{-2}$,
$\bar a_1 = (-1 \ldots +1) \times 10^{-3} M_\pi^{-2}$.
We find however (in agreement with~\cite{borasoy})
that the effects of the P-wave $\bar a_1$ are absolutely negligible
(about two orders of magnitude smaller than the changes induced by the variation in the $\pi\eta$ S-wave
shown in the following), and therefore set $\bar a_1=0$ in the sequel.
\end{sloppypar}

There is no very precise information on the $\eta'\to\eta\pi\pi$ Dalitz plot parameters.
We will use the central values (without errors) of
the most recent determinations 
done by the VES collaboration for the charged channel~\cite{VES}, which, 
adjusting the normalization to the neutral decay, 
read $a=-0.133$, $b=-0.116$, $d=-0.094$.  
We have checked using rather different older data on the neutral channel
($a=-0.116$, $b=d=0$~\cite{Alde}) that, while the overall Dalitz plot distribution of course
looks very different, all statements about the cusp behavior hold in exactly the same way.
We only discuss normalized decay spectra, setting $\N=1$.

\begin{figure}
 \centering
 \includegraphics[width=\linewidth]{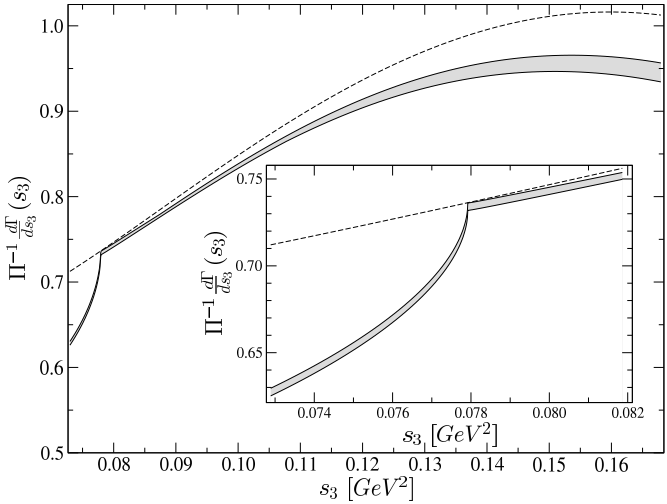}
 \caption{The decay rate $d\Gamma/ds_3$ divided by phase space. 
 The Dalitz plot parameters have been matched to the tree-level amplitude. 
 The dashed line is the tree result, the gray band shows the full result
 under variation of the $\pi\eta$ threshold parameters.
 The insert focuses on the cusp region around the $\pi^+\pi^-$ threshold.
 \label{fig:Twoloopw/oren}}
\end{figure}

\begin{sloppypar}
The decay spectrum $d\Gamma/ds_3$ for $\eta'\to\eta\pi^0\pi^0$ is calculated by
\begin{align}
\frac{d\Gamma}{ds_3} &= \frac{1}{512\pi^3 M_{\eta'}^3} \int_{s_1^-(s_3)}^{s_1^+(s_3)} ds_1 |\M_N(s_1,s_3)|^2 ~, \\
s_1^\pm(s_3) &= \me+\mpn+\frac{1}{2}\Bigl\{\mep-\me-s_3 \nn
& \quad \pm \frac{1}{s_3}\lambda^{1/2}(\mep,\me,s_3)\lambda^{1/2}(s_3,\mpn,\mpn) \Bigr\} ~, \nonumber
\end{align}
which we normalize by the phase space factor
\beq
\Pi(s_3) = \frac{s_1^+(s_3)-s_1^-(s_3)}{512\pi^3 M_{\eta'}^3} ~.
\eeq
In Fig.~\ref{fig:Twoloopw/oren} 
we display the normalized decay spectrum $\Pi^{-1}(s_3)d\Gamma/ds_3$
for $\eta'\to\eta\pi^0\pi^0$, with the coupling constants $G_i$, $H_i$ matched to the 
VES Dalitz plot parameters directly according to~\eqref{NCTreematching}.
The cusp effect is prominently visible below the charged-pion threshold and  amounts 
to a reduction of the decay spectrum up to around $10\%$.
Furthermore, the final-state interaction reduces the spectrum for large $s_3$ by about $5\%$,
largely due to $\pi\pi$ rescattering.  
The effects generated by varying $\pi\eta$ threshold parameters (as seen by the gray band 
in Fig.~\ref{fig:Twoloopw/oren}) are moderate, but particularly small in the cusp region.
We remark that the sign of the cusp is fixed by (approximate) isospin symmetry, $H_i = -\sqrt{2} G_i$.
\end{sloppypar}

It is obvious from Fig.~\ref{fig:Twoloopw/oren} that the amplitude does not reproduce the
experimental Dalitz plot parameters any more -- they are renormalized by the final-state interactions.
We therefore re-adjust the tree-level couplings such that the \emph{full} amplitude squared~\eqref{eq:DalitzDef}
yields the VES parameters.  The renormalized tree-level parameters, according to~\eqref{NCTreematching}, 
e.g.\ correspond to 
$\N_{\rm ren}=1.015~(1.021)$, 
$a_{\rm ren}=-0.168 ~(-0.181)$, 
$b_{\rm ren}=-0.108~(-0.106)$, 
$d_{\rm ren}=-0.099~(-0.097)$,
for $\bar a_0 = 0.016$, $\bar b_0 = 0.010\,M_\pi^{-2}$ ($\bar a_0 = 0$, $\bar b_0 = 0$).
\begin{figure}
 \centering
 \includegraphics[width=\linewidth]{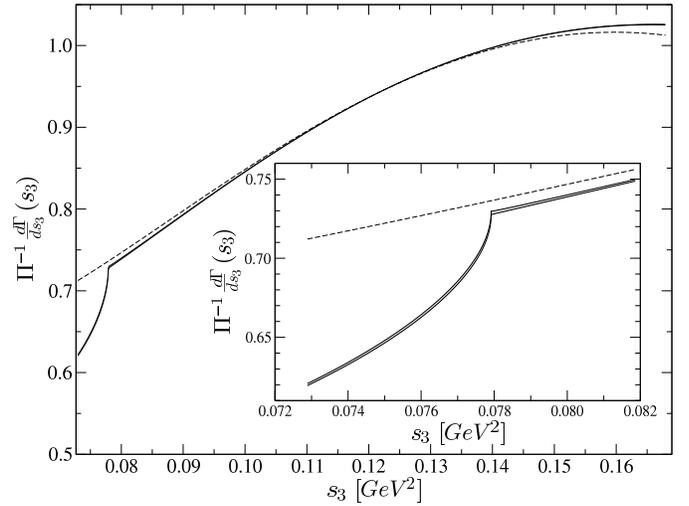}
 \caption{The decay rate ${d\Gamma}/{ds_3}$ divided by phase space. 
The Dalitz plot parameters have been matched to the full amplitude. 
We show tree (dashed) and full result (solid), the latter again for varying $\pi\eta$ threshold parameters.
Effects of these parameters are observed to be significantly reduced compared to Fig.~\ref{fig:Twoloopw/oren}.
The insert magnifies the $\pi^+\pi^-$ threshold region.}
 \label{fig:Twoloopwren}
\end{figure}
The decay spectra resulting from this procedure are shown in Fig.~\ref{fig:Twoloopwren}.  
We see that the full result follows the tree-level spectrum closely except for small deviations
close to the kinematic limits, and the prominent cusp below the $\pi^+\pi^-$ threshold.
The uncertainty band due to $\pi\eta$ scattering 
has shrunk to a very narrow line: the effects of the third-particle rescattering
can be absorbed to a large extent in a redefinition of the polynomial part.  

Integrating the spectrum in the region $4\mpn\leq s_3\leq4\mpi$, we find that the cusp reduces the number
of events in that region with respect to the tree distribution (or no $\pi\pi$ rescattering) by more
than 8\%, compared to about 13\% in $K^+\to\pi^0\pi^0\pi^+$ (see e.g.~\cite{MadigozhinCapri}),
or less than 2\% in $\eta\to 3\pi^0$~\cite{Gullstrom}.

\begin{figure}
 \centering
 \includegraphics[width=\linewidth]{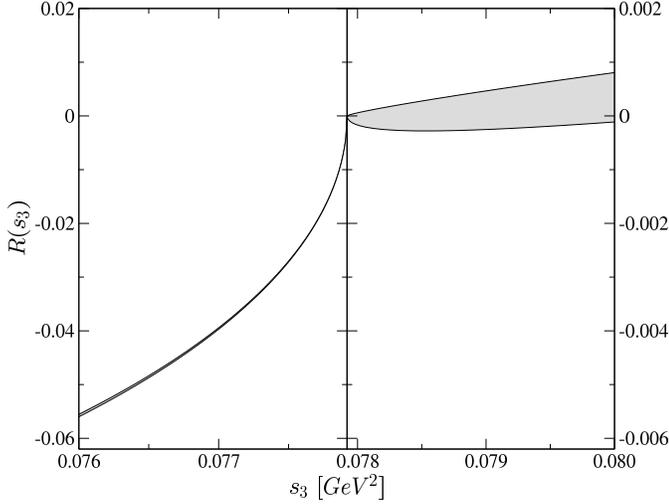} 
 \caption{Difference between full and tree decay rate ${d\Gamma}/{ds_3}$
in the cusp region, divided by the phase space, and shifted to 0 at $s_3=4\mpi$.
Note that the scale above the cusp has been increased by a factor of 10.
 \label{fig:Twoloopcusp}} 
\end{figure}
In Fig.~\ref{fig:Twoloopcusp} we zoom in further into the cusp region to
investigate the expected two-loop cusp above threshold.
We plot the decay spectrum, with the tree spectrum subtracted, shifted to 0 at threshold,
\begin{align}
R(s_3) &= \Pi^{-1}(s_3) \biggl[ \frac{d\Gamma_\textrm{full}}{ds_3} -  \frac{d\Gamma_\textrm{tree}}{ds_3} \biggr] \nn
& \quad - \Pi^{-1}(4\mpi)\biggl[\frac{d\Gamma_\textrm{full}}{ds_3} - \frac{d\Gamma_\textrm{tree}}{ds_3}\biggr]_{s_3=4\mpi} 
~.
\end{align}
We increase the scale above threshold by a factor of 10.
Obviously the two-loop cusp (above threshold) is highly suppressed compared to the one-loop cusp (below threshold). 
One would  expect the two-loop cusp effect to be smaller than the one generated by one-loop diagrams,
as it is suppressed by another power in $\pi\pi$ or $\pi\eta$ threshold parameters; 
however, a suppression by more than two orders of magnitude may seem surprising at first.
The explanation for this observation can be found resorting to the 
\emph{threshold theorem}~\cite{Cabibbo,CGKR,Photons}: 
in the direct vicinity of the $\pi^+\pi^-$ threshold, one can parameterize the amplitude (up to three loops,
in the absence of photons) as
\begin{align}
\M_N &= \alpha_0+i\bigl( \alpha_1+\alpha_1'q_{+-}(s_3)\bigr)+\bigl(\alpha_2-\alpha_2'q_{+-}(s_3)\bigr) \nn
& \quad +i\bigl( \alpha_3+\alpha_3'q_{+-}(s_3)\bigr) +\Order(a^4)~,
\label{eq:MNthresh}
\end{align}
where the real parameters $\alpha_i$, $\alpha_i'$ are of $\Order(a^i)$. This leads to
\begin{align}
\left|\M_N\right|^2 &=\textrm{reg.}-2 \bigl(\alpha_0(\alpha_1'+\alpha_3')+\alpha_2\alpha_1'+\alpha_1\alpha_2'\bigr)
\nn & \quad \times \sqrt{-q_{+-}^2(s_3)} +\Order(a^5) \label{eq:cuspbelow}
\end{align}
below threshold, where ``reg.''\ denotes polynomial (non-singular) terms in the vicinity of $s_3=4\mpi$, and
\beq
\left|\M_N\right|^2=\textrm{reg.}-2 \bigl(\alpha_0\alpha_2'-\alpha_1\alpha_1'\bigr)q_{+-}(s_3)+\Order(a^4) 
\label{eq:cuspabove}
\eeq
above threshold. 
The threshold theorem states that the coefficients $\alpha_i'$ are proportional to $\M_C$ 
in the appropriate kinematics,
\beq\label{threstheorem}
\M_C\bigl(s_1=s_2,s_3=4\mpi\bigr) \propto \alpha_1'+i\alpha_2'+\alpha_3' + \Order(a^3)~,
\eeq 
where the factor of proportionality includes the $\pi\pi$ charge- \eject\noindent
exchange scattering length, while obviously
\beq
\M_N\bigl(s_1=s_2,s_3=4\mpi\bigr)= \alpha_0+i\alpha_1+ \alpha_2 + \Order(a^3)~.
\eeq
In the isospin limit $ \alpha_i\propto-\sqrt{2}\alpha_{i+1}'$ and,
according to~\eqref{eq:cuspabove}, the two-loop effects on the cusp would exactly cancel. 
Numerically, we find with our set of parameters
\begin{align}
\M_C\bigl(s_1=s_2,s_3=4\mpi\bigr) &= -1.22\,(\textrm{tree})-0.076i\,(\textrm{1-loop})\nn&\quad+0.0046\,(\textrm{2-loop}) ~,\nn
\M_N\bigl(s_1=s_2,s_3=4\mpi\bigr) &= 0.86\,(\textrm{tree})+0.050i\,(\textrm{1-loop})\nn&\quad-0.0039\,(\textrm{2-loop}) ~,
\end{align}
which leads to a suppression of the two-loop cusp by about a factor of 250.
Incidentally, \eqref{eq:cuspbelow} also allows to estimate the effect of three-loop (or $\Order(a^3)$)
contributions to the cusp, which do \emph{not} vanish in the isospin limit.
We find that $\Order(a^3)$ terms ought to reduce the leading $\Order(a)$ cusp by about 0.5\%.
Therefore we conclude that, unlike in $K^+\to\pi^0\pi^0\pi^+$ decays, where two-loop contributions
are an essential ingredient to the proper theoretical description of the amplitude in the threshold
region, the cusp in $\eta'\to\eta\pi^0\pi^0$ is entirely dominated by the leading $\Order(a)$ 
rescattering effects.

\begin{figure}
 \centering
 \includegraphics[width=0.965\linewidth]{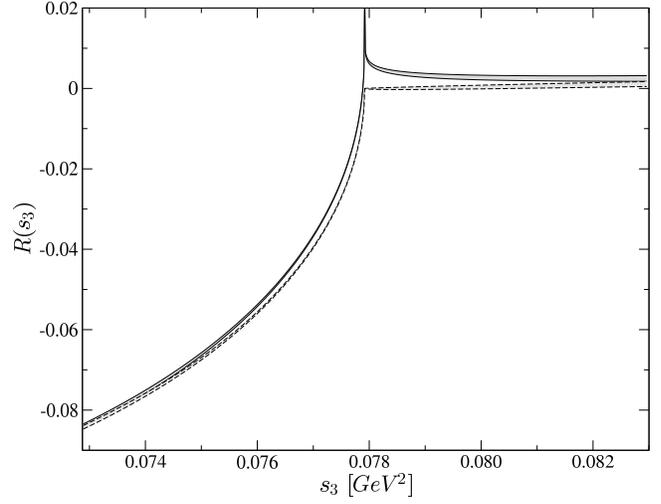}
 \caption{Difference between full and tree decay rate ${d\Gamma}/{ds_3}$
 in the cusp region, divided by the phase space.  
 The dashed band is without, the full band with radiative corrections.
 The latter displays the logarithmic singularity at threshold; compare~\eqref{eq:replacementCoulomb}.
 \label{fig:cusprad}}
\end{figure}
\begin{sloppypar}
Finally, we briefly comment on the effect of the radiative corrections discussed in Sect.~\ref{sec:rad}.  
Figure~\ref{fig:cusprad} shows the cusp region similarly to Fig.~\ref{fig:Twoloopcusp}, 
with and without the replacement~\eqref{eq:replacementCoulomb}.  
While the logarithmic singularity at threshold is visible, the effect becomes very small
away from $s_3=4\mpi$.  These corrections clearly only matter in experimental analyses
with very high resolution and statistics.
\end{sloppypar}

\eject

\section{Summary and conclusion}\label{sec:summary}

\begin{sloppypar}
In this article, we have generalized the formalism of non-relativistic effective field theory
to describe the analytic structure, and in particular the cusp effect, 
in $\eta'\to\eta\pi\pi$ decays.  
We have shown how to construct an effective Lagrangian that reproduces 
the next-to-leading threshold parameters (effective range, P-wave scattering length)
in $\pi\eta$ scattering, and derived the decay amplitudes
up to two loops, including $\Order(\epsilon^4)$, $\Order(a\epsilon^5)$, 
$\Order(a_{\pi\pi}^2\epsilon^6,a_{\pi\pi}a_{\pi\eta}\epsilon^2,a_{\pi\eta}^2\epsilon^2)$.
These amplitudes are the central result of our investigation, and ought to be 
employed in future precision studies of the $\eta'\to\eta\pi\pi$ Dalitz plot.
\end{sloppypar}

Invoking theoretical information on the coupling constants involved, we have also \emph{predicted}
the size of the cusp effect, and shown that it reduces the decay spectrum below 
the charged-pion threshold by more than 8\%.  This is a much more sizeable effect
than e.g.\ in $\eta\to3\pi^0$ decays.  Approximate isospin symmetry dictates
that the cusp of $\Order(a^2)$ above threshold is strongly suppressed, 
and three-loop effects can be estimated to yield a correction below 1\%.
Therefore the threshold singularity in $\eta'\to\eta\pi^0\pi^0$ is determined
to very high precision by the leading $\Order(a)$ rescattering effect.
Experimental verifications of these predictions at various 
laboratories~\cite{Beck,MAMI-C1,MAMI-C2,WasaAdam,WASA,KLOElett,KLOE,BES} are eagerly  awaited.

\begin{acknowledgement}
\begin{sloppypar}
\textit{Acknowledgements.}
We would like to thank Akaki Rusetsky for useful discussions,
and Martin Hoferichter for comments on the manuscript.
Partial financial support by the Helmholtz Association through funds provided
to the virtual institute ``Spin and strong QCD'' (VH-VI-231), 
by the European Community-Research Infrastructure Integrating Activity 
``Study of Strongly Interacting Matter''
(acronym HadronPhysics2, Grant Agreement n.~227431) under the Seventh 
Framework Programme of the EU,
and by DFG (SFB/TR 16, ``Subnuclear Structure of Matter'') is gratefully
acknowledged. 
\end{sloppypar}
\end{acknowledgement}


\begin{appendix}

\renewcommand{\theequation}{\thesection.\arabic{equation}}

\setcounter{equation}{0}
\section{The two-loop function}\label{twoloopfunc}

The analytic representation of the two-loop formula is given as \cite{BFGKR}
\beq
F_k(M_a,M_b,M_c,M_d,s_k)=\bar\N(2Af_1+Bf_0)+\Order(\epsilon^4)~,
\eeq
with
{\allowdisplaybreaks
\begin{align}
\bar\N&=\frac{1}{256\pi^3\sqrt{s_k}}\frac{\lambda^{1/2}(s_0,M_a^2,M_b^2)}{s_0\sqrt{\Delta^2-\frac{(1+\delta)^2}{4}{\bf Q}_k^2}}~,\nn
f_0&=4\left(v_1+v_2-\bar v_2+h\right)~,\nn
f_1&=\frac{4}{3}\left(y_1(v_1-1)+y_2(v_2-1)-\bar y_2(\bar v_2-1)+h\right)~,\nn
h&=\frac{1}{2}\log\left(\frac{1+{\bf Q}_k^2/s_k}{1+\bar {\bf Q}_k^2/\bar s_k}\right)~,~
\bar {\bf Q}_k^2={\bf Q}_k^2(\bar s_k)~,\nn
v_i&=\sqrt{-y_i}\arctan\frac{1}{\sqrt{-y_i}}~,~i=1,2~;\nn
\bar v_2&=\sqrt{-\bar y_2}\arctan{\frac{1}{\sqrt{-\bar y_2}}}~,~
\bar y_2=y_2(\bar s_k)~, \nn
\bar s_k&=(M_c+M_d)^2 ~,\nn
y_{1,2}&=\frac{-B\mp \sqrt{B^2-4AC}}{2A}~,~
A=-\frac{{\bf Q}_k^2}{s_k}(M_c^2+\Delta^2)~,\nn
B&=q_{cd}^2(s_k)-\Delta^2+\frac{{\bf Q}_k^2}{s_k}M_c^2~,~
C=-q_{cd}^2(s_k)~,\nn
s_0&=M_{\eta'}^2+M_c^2-2M_{\eta'}\left(M_c^2+\frac{{\bf Q}_k^2(1+\delta)^2}{4}\right)^{1/2}~,\nn
\Delta^2&=\frac{\lambda(M_{\eta'}^2,M_c^2,(M_a+M_b)^2)}{4M_{\eta'}^2}~,~\delta=\frac{M_c^2-M_d^2}{s_k}~.
\end{align}}

\setcounter{equation}{0}
\section{\boldmath{$\pi\eta$} threshold parameters}\label{pietathresh}

\begin{sloppypar}
$\pi\eta$ scattering has been calculated up to $\Order(p^4)$ in chiral perturbation theory (ChPT)
in~\cite{BKM:pieta}.  We use the form quoted in~\cite{GP} for the expansion near threshold.
The scattering lengths $a_0^{\pi\eta}$ and $a_1^{\pi\eta}$ are discussed extensively in~\cite{Kolesar}.
\end{sloppypar}

The S-wave $\pi\eta$ scattering length to $\Order(p^4)$ is given by\footnote{All $\pi\eta$ threshold
parameters are given in the isospin limit.  Note that isospin breaking
in $\pi\eta$ at tree level only affects the scattering length at $\Order((m_u-m_d)^2)$ and is therefore negligible; 
there are no electromagnetic effects at that order.}
\begin{align}
\bar a_0 &= \frac{\mpi}{96\pi F_\pi^2} \biggl\{ 1+\frac{96}{F_\pi^2} \biggl[ \Bigl(L_1^r+L_2^r+\frac{L_3}{2}-L_4^r-\frac{L_5^r}{6}+L_6^r\nn 
& - L_7\Bigr)\me + \Bigl(L_7+\frac{L_8^r}{2}\Bigr)\mpi \biggr]-\frac{1}{16\pi^2F_\pi^2}\biggl[ 3\mpi \logpi \nn
&+ \bigl(15\me+\mpi\bigr)\logk + \Bigl(\frac{4}{3}\me+\mpi\Bigr)\loge \nn
&- \frac{4M_\pi^4}{3(\me\!-\!\mpi)}\log\frac{M_\pi}{M_\eta} +  A_+ + A_-\! - \frac{25\me\!+\!7\mpi}{3} \biggr]\!
\biggr\} , \nn
A_\pm &= \bigl(3M_\eta\pm M_\pi\bigr)^2 \frac{\sqrt{2M_\eta(M_\eta \mp M_\pi)}}{M_\eta \pm M_\pi} \nn
& \qquad \times\arctan \biggl(\frac{M_\eta \pm M_\pi}{\sqrt{2M_\eta(M_\eta \mp M_\pi)}}\biggr)  ~, \label{eq:a0} 
\end{align}
where we have made use of the Gell-Mann--Okubo relation $4\mk=3\me+\mpi$ in the $\Order(p^4)$ corrections.
$\bar a_0$ can be related to the $I=2$ $\pi\pi$ scattering length
\begin{align}
a_0^2 &= -\frac{\mpi}{16\pi F_\pi^2} \biggl\{1-\frac{32\mpi}{F_\pi^2}
\biggl[L_1^r+L_2^r+\frac{L_3}{2}-L_4^r-\frac{L_5^r}{2} \nn
& +L_6^r+\frac{L_8}{2} \biggr] + \frac{\mpi}{16\pi^2 F_\pi^2}\biggl[ 3\logpi + \frac{1}{9}\loge - \frac{4}{9} \biggr] \biggr\} ,
\end{align}
as well as $\Delta_F = F_K/F_\pi-1$ and 
$\Delta_{\rm GMO}= (4\mk-3\me-\mpi)/(\me-\mpi)$~\cite{GL:SU3}
in the form of the low-energy theorem
\begin{align}
\bar a_0 &= \frac{\mpi}{96\pi F_\pi^2} \biggl\{ 1+ \frac{48\pi F_\pi^2 \me}{M_\pi^4} \Delta a_0^2
+ \frac{8}{3}\frac{3\me+\mpi}{\me-\mpi} \Delta_F \nn
&+ \frac{4}{3} \Delta_{\rm GMO} +\frac{1}{16\pi^2F_\pi^2} \biggl[ 
\frac{9M_\eta^4-6\me\mpi+M_\pi^4}{3(\me-\mpi)} \log\frac{M_\pi}{M_\eta}\nn
&+ 2 \frac{3M_\eta^4-10\me\mpi-M_\pi^4}{\me-\mpi}\log\frac{M_\pi}{M_K} - A_+ - A_-  \nn
& +7\biggl(\me +\frac{\mpi}{3}\biggr)\biggr] \biggr\} ~, \label{eq:a0LET}
\end{align}
where $\Delta a_0^2 = a_0^2 + \mpi/(16\pi F_\pi^2)$.
We find the (modified) $\pi\eta$ effective range $\bar b_0$ 
and P-wave scattering length $\bar a_1$
at $\Order(p^4)$ 
\begin{align}
\bar b_0 &= \frac{1}{\pi F_\pi^4} \biggl\{\biggl(L_1^r+L_2^r+\frac{L_3}{2}-\frac{L_4^r}{2}\biggr)
\bigl(\me+\mpi\bigr) \nn
&+ \biggl( L_2^r+\frac{L_3}{3}\biggr) M_\pi M_\eta \biggr\} 
+\frac{1}{512\pi^3\Fpi^4} \biggl\{ \frac{4}{3}\mpi\log\frac{M_\pi}{M_K} \nn 
&+\frac{2M_\pi^4(3\me-2M_\eta M_\pi+3\mpi)}{9(M_\eta-M_\pi)^3(M_\eta+M_\pi)}\log\frac{M_\pi}{M_\eta} -\frac{\tilde A_+}{6}\nn
&-7\biggl(\me\!+\!\frac{6}{7}M_\pi M_\eta\!+\!\mpi\biggr)\logk 
 -\frac{(\me+\mpi)\tilde A_-}{6(M_\eta-M_\pi)^2}\nn
& +\Bigl[135M_\eta^6-108M_\eta^5M_\pi+107M_\eta^4\mpi-214M_\eta^3M_\pi^3\nn
&+ 177\me M_\pi^4 \!-\! 2M_\eta M_\pi^5 \!+\! M_\pi^6\Bigr]
\bigl(54\me(M_\eta\!-\!M_\pi)^2\bigr)^{-1} \! \biggr\} , \nn
\bar a_1 &= -\frac{1}{3\pi F_\pi^4} 
\biggl\{\biggl(L_1^r+\frac{L_3}{6}-\frac{L_4^r}{2}\biggr)\bigl(\me+\mpi\bigr) \nn
&-\biggl(L_2^r+\frac{L_3}{3}\biggr)M_\eta M_\pi\biggr\} 
+ \frac{1}{4608\pi^3 F_\pi^4} \biggl\{ 4\mpi\logpi\nn
&-\frac{2M_\pi^4(M_\eta+M_\pi)}{3(M_\eta-M_\pi)^3}\log\frac{M_\pi}{M_\eta} +\frac{M_\eta M_\pi \tilde A_-}{(M_\eta-M_\pi)^2} \nn
&-\Bigl(3\me-18M_\pi M_\eta+7\mpi\Bigr)\logk  \nn
& - \Bigl[27M_\eta^6+108M_\eta^5M_\pi-215M_\eta^4\mpi+214M_\eta^3M_\pi^3\nn
&- 39\me M_\pi^4\!+\!2M_\eta M_\pi^5 \!-\! M_\pi^6\Bigr]
\bigl(18\me(M_\eta\!-\!M_\pi)^2\bigr)^{-1}  \biggr\} , \nn
\tilde A_\pm &= \frac{12M_\eta^4\pm3M_\eta^3M_\pi-\!15M_\eta\mpi(M_\eta\pm M_\pi)-M_\pi^4}
{2\me(3M_\eta\pm M_\pi)(M_\eta\mp M_\pi)} \,A_\pm . 
\end{align}
As the $\pi\pi$ scattering amplitude in SU(3) only depends on the linear combination
$2L_1^r+L_3$~\cite{GP}, it is obvious that one cannot formulate low-energy theorems
for $\bar b_0$ and $\bar a_1$ in terms of $\pi\pi$ threshold parameters
(or $\pi K$ scattering lengths~\cite{Kubis:piK2}).  
We therefore refrain from recasting these results in alternative forms.

We use two different sets of $\Order(p^4)$ low-energy constants for numerical evaluation~\cite{BEG:Daphne,ABT}.
A major difficulty consists in estimating the combined errors, as the various
uncertainties for the low-energy constants are strongly correlated.
The errors quoted in Table~\ref{tab:Nab} are obtained by naive error propagation, neglecting
any correlations; we consider the uncertainties thus obtained significantly overestimated.
In the case of $\bar a_0$, we also use the low-energy theorem~\eqref{eq:a0LET}, 
with $\Delta a_0^2 = 0.0012\pm0.0010$~\cite{CGL}, $\Delta_{\rm GMO} = 0.196$, and $\Delta_F = 0.193 \pm 0.006$~\cite{pdg}.
Table~\ref{tab:Nab} shows the results for $\bar a_0$, $\bar b_0$, $\bar a_1$ thus obtained; 
in the cases of $\bar a_0$ and $\bar a_1$, these are consistent with the findings in~\cite{Kolesar}.
\begin{table}
\centering
\renewcommand{\arraystretch}{1.3}
\begin{tabular}{ccrclrclrcl}
\hline
&  \!CA\! & \mc{Ref.~\cite{BEG:Daphne}} & \mc{Ref.~\cite{ABT}} & \mc{LET} \\
\hline
  $10^3        \,\bar a_0 $ & \!7.6\! &\!$15.7$&\tpm&\tbk$23.9$\! & $ 9.8$&\tpm&\tbk$15.8$ & $-0.2$&\tpm&\tbk$7.7$\\
\!$10^3 M_\pi^2 \,\bar b_0 $ &   0     &   $9.9$&\tpm&\tbk$22.9$\! &  $0.4$&\tpm&\tbk$18.8$ & \mc{~~---}\\
\!$10^3 M_\pi^2 \,\bar a_1 $ &   0     &   $0.9$&\tpm&\tbk$ 3.7$   & $-0.8$&\tpm&\tbk$ 2.3$ & \mc{~~---}\\
\hline
& & \mc{$(\pi\pi)_{00}$} & \mc{$(\pi\pi)_x$} & \mc{$(\pi\pi)_{+-}$} \\
\hline
$10^3        \,a_0 $   & & $40.9$&\tpm&\tbk$1.7$ &  $ -90.0$&\tpm&\tbk$1.7$& $70.2$&\tpm&\tbk$1.7$  \\
\!$10^3 M_\pi^2 \,b_0 $ & & $38.5$&\tpm&\tbk$2.0$ &\!$-118.8$&\tpm&\tbk$2.0$& $78.6$&\tpm&\tbk$2.0$ \\
\!$10^3 M_\pi^2 \,a_1 $ & & \mc{---}              & \mc{---}                & $19.0$&\tpm&\tbk$0.3$ \\
\hline
\end{tabular}
\renewcommand{\arraystretch}{1.0}
\caption{Numerical results for the $\pi\eta$ threshold parameters.\label{tab:Nab}
The first column ``CA'' refers to the $\Order(p^2)$ values, 
the second and third column show the $\Order(p^4)$ results, evaluated
with the low-energy constants taken from~\cite{BEG:Daphne,ABT}.
Errors are obtained naively by adding individual uncertainties in quadrature.
The lower three columns show the $\pi\pi$ threshold parameters for the physical channels
elastic $\pi^0\pi^0$ scattering $(\pi\pi)_{00}$, charge exchange $(\pi\pi)_x$, and elastic $\pi^+\pi^-$ $(\pi\pi)_{+-}$.
For details, see text.}
\end{table}
The conclusion is that chiral symmetry does not make very precise predictions for the $\pi\eta$ threshold parameters.
The next-to-leading order corrections for $\bar a_0$ can be as large as the current-algebra value, 
the magnitude of the effective range is very badly constrained, and
not even the sign is fixed for $\bar a_1$.
For numerical evaluation in the main text, we decide to vary the $\pi\eta$ threshold parameters in the ranges
$\bar a_0 = (0 \ldots +16)\times 10^{-3}$, 
$\bar b_0 = (0 \ldots +10) \times 10^{-3} M_\pi^{-2}$,
$\bar a_1 = (-1 \ldots +1) \times 10^{-3} M_\pi^{-2}$.
We consider these reasonable, although not the most conservative limits possible.  
They comprise the values given in~\cite{BKM:pieta} and most of the parameter ranges discussed
in~\cite{Kolesar}.

To put these numbers into perspective,
Table~\ref{tab:Nab} also shows the $\pi\pi$ threshold parameters~\cite{CGL} for the different
physical channels (corrected for tree-level isospin breaking in the S-wave scattering lengths~\cite{KnechtUrech}),
in the same units.
The errors are approximate only and, in case of the S-waves, propagated from the dominant $I=0$ threshold parameters.
The comparison demonstrates that, even within
a large uncertainty range, $\pi\eta$ scattering in general is much weaker than $\pi\pi$ scattering
and should therefore have far less influence on the decay properties of $\eta'\to\eta\pi\pi$
via final-state interactions.
We furthermore remark that
a simple unitarization model~\cite{Oller} (that reproduces the $a_0(980)$ resonance within reasonable accuracy)
indicates that the $\pi\eta$ phase stays below $5^\circ$ 
in the range up to $\sqrt{s} \leq M_{\eta'}-M_{\pi^0}\approx 823$~MeV, therefore the influence
of the $a_0(980)$ resonance is not yet severe and the effective range expansion still applicable.

\end{appendix}


\end{document}